\documentclass[%
 reprint,
 amsmath,amssymb,
 aps,
]{revtex4-2}

\usepackage{graphicx}
\usepackage{dcolumn}
\usepackage{bm}
\usepackage{xcolor}
\usepackage{comment}
\usepackage{soul}

\begin{document}

\preprint{APS/123-QED}

\title{Universal bounds on entropy production inferred from observed statistics}

\author{Eden Nitzan}
 \affiliation{School of Physics \& Astronomy, Faculty of Exact Sciences, Tel Aviv University, Tel Aviv 6997801, Israel}
 \author{Aishani Ghosal}
  \affiliation{Department of Biomedical Engineering, Tel Aviv University, Tel Aviv 6997801, Israel}
\author{Gili Bisker}%
 \email{bisker@tauex.tau.ac.il}
 \affiliation{%
 Department of Biomedical Engineering, Faculty of Engineering\\
 The Center for Physics and Chemistry of Living Systems\\ The Center for Nanoscience and Nanotechnology\\ The Center for Light Matter Interaction\\
 Tel Aviv University, Tel Aviv 6997801, Israel
}

\date{\today}

\begin{abstract}
Nonequilibrium processes break time-reversal symmetry and generate entropy. Living systems are driven out-of-equilibrium at the microscopic level of molecular motors that exploit chemical potential gradients to transduce free energy to mechanical work, while dissipating energy. The amount of energy dissipation, or the entropy production rate (EPR),  sets thermodynamic constraints on cellular processes. Practically, calculating the total EPR in experimental systems is challenging due to the limited spatiotemporal resolution and the lack of complete information on every degree of freedom. 
Here, we propose a new inference approach for a tight lower bound on the total EPR given partial information, based on an optimization scheme that uses the observed transitions and waiting times statistics. We introduce hierarchical bounds relying on the first- and second-order transitions, and the moments of the observed waiting time distributions, and apply our approach to two generic systems of a hidden network and a molecular motor, with lumped states. Finally, we show that a lower bound on the total EPR can be obtained even when assuming a simpler network topology of the full system.
\end{abstract}

\maketitle

\section{\label{sec:introduction}Introduction\protect}

Advances in experimental techniques over the last few decades have opened new possibilities for studying systems at the single-molecule level \cite{bustamante2021optical, kinz2021bayesian,bustamante2020single}. In parallel, new theoretical approaches of stochastic thermodynamics for studying the physics of nonequilibrium, small fluctuating systems have emerged \cite{bustamante2005unfolding, seifert2012stochastic,van2015ensemble}. These include the mathematical relations describing symmetry properties of the stochastic quantities like work \cite{van2003stationary, douarche2006work, sabhapandit2012heat} heat \cite{sabhapandit2012heat, visco2006work}, and entropy production \cite{wang2002experimental, ciliberto2013heat}, 
leading to fundamental limits on physical systems like heat engines \cite{martinez2016brownian, van2012efficiency, verley2014unlikely} refrigerators \cite{mohanta2022universal}, and biological processes \cite{bo2015thermodynamic, saadat2020thermodynamic}.

Living systems operate far-from-equilibrium and constantly produce entropy. At the molecular level, the hydrolysis of fuel molecules, such as Adenosine triphosphate (ATP), powers nonequilibrium cellular processes, utilizing part of the liberated free energy for physical work, while the rest is dissipated \cite{seifert2012stochastic}.
The dissipation, or entropy production, is a signature of irreversible processes and can be used as a direct measure of the deviation from thermal equilibrium \cite{li2019quantifying, fodor2016far, maes2003time, parrondo2009entropy}.
Therefore, the entropy production rate plays an important role in our understanding of the physics, and underlying mechanism, governing biological and chemical processes \cite{bo2015thermodynamic, saadat2020thermodynamic,martinez2016brownian, van2012efficiency, verley2014unlikely, pietzonka2016universal}.

Various studies have focused on estimating the mean entropy production rate using the thermodynamic uncertainty relations (TUR) using current fluctuations \cite{shiraishi2021optimal,horowitz2020thermodynamic, gingrich2016dissipation, barato2015thermodynamic,manikandan2021quantitative,manikandan2020inferring}, fluctuations of first passage time \cite{PhysRevLett.119.170601, PhysRevResearch.3.L032034}, 
kinetic uncertainty relation in terms of the activity \cite{di2018kinetic}, or unified thermodynamic and kinetic uncertainty relations \cite{Vo_2022}. Other approaches utilize waiting-time distributions \cite{skinner2021estimating, D2CP03064K}, machine learning \cite{otsubo2020estimating, kim2020learning, bae2022inferring}, and single trajectory data \cite{roldan2010estimating, otsubo2022estimating, lander2012noninvasive}. Additional studies calculate higher moments of the full probability density function of the entropy production \cite{padmanabha2022fluctuations}, use irreversible currents in stochastic dynamics described by a set of Langevin equations \cite{dechant2018entropic}, or linear response theory \cite{pietzonka2016universal}. 

Estimating the total EPR is only possible if we have knowledge regarding all of the degrees of freedom that are out-of-equilibrium \cite{PhysRevLett.98.080602}. However, due to practical limitations on the spatiotemporal resolution, not all of them can be experimentally accessible, and one can only obtain a lower bound on the total EPR for partially observed or coarse-grained systems \cite{bisker2017hierarchical}.

The passive partial entropy production rate, $\sigma_{\text{pp}}$, is an estimator for the EPR calculated from the transitions between two observed states, which bounds the total EPR \cite{bisker2017hierarchical, shiraishi2015role, shiraishi2015fluctuation, polettini2017effective}. This estimator, however, fails to provide a non-zero bound in case of vanishing current over the observed link, \emph{i.e.}, at stalling conditions \cite{bisker2017hierarchical}.
Other EPR estimators for partially observed systems based on inequality relations like the TUR \cite{shiraishi2021optimal, horowitz2020thermodynamic, gingrich2016dissipation, di2018kinetic, vo2022unified} also fail to provide a non-trivial bound on the total EPR in the absence of net flux in the system.

The Kullback-Leibler Divergence (KLD) estimator, $\sigma_{\text{KLD}}$, is based on the KLD, or the relative entropy, between the time-forward and the time-revered path probabilities \cite{kawai2007dissipation, maes1999fluctuation, roldan2021quantifying, horowitz2009illustrative, gaveau2014dissipation, gaveau2014relative, maes2003time}.
For semi-Markov processes, this estimator is a sum of two contributions. The first stems from transitions irreversibility or cycle affinities, $\sigma_{\text{aff}}$, whereas the second stems 
from broken time-reversal symmetry reflected in  irreversibility in waiting time distributions (WTD),  $\sigma_{\text{WTD}}$ \cite{martinez2019inferring}. Using the KLD estimator, one can obtain a non-trivial lower bound on the total EPR for second-order semi-Markov processes even in the absence of the net current \cite{martinez2019inferring, D2CP03064K, van2022thermodynamic, hartich2021comment, bisker2022comment}. 
Moreover, a lower bound on the total EPR can be obtained from the KLD between transition-based WTD \cite{van2022thermodynamic, roldan2021quantifying,van2022time}.

Recently developed estimators solved an optimization problem to obtain a lower bound on the entropy production. For a discrete-time model, Ehrich proposed to search over the possible underlying systems that maintain the same observed statistics using knowledge on the number of hidden states \cite{ehrich2021tightest}. For continuous-time models, Skinner and Dunkel minimized the EPR on a canonical form of the system that preserved the first- and second-order transition statistics to yield a lower bound on the total EPR, $\sigma_2$ \cite{skinner2021improved}. The authors also formulated an optimization problem to infer the EPR in a system with two observed states using the waiting time statistics \cite{skinner2021estimating}.

In this paper, we provide a tight bound on the total EPR by formulating an optimization problem based on the statistics of both transitions and waiting times. We use the first- and second-order statistics for the mass transition rates, and any chosen number of moments of the observed waiting time distributions. 
For a system with a known topology, we calculate the analytical expressions of the statistics as functions of the mass rates and the steady-state probabilities, which describe a possible underlying system and are used as variables in the optimization problem. These analytical expressions are then used to constrain the optimization variables to match the observed statistics.
We show for a few continuous-time Markov chain systems that using the constraints of the mass rates and only the first moment of the WTD already provides close-to-total EPR value. Our approach outperforms other estimators, such as $\sigma_{\text{pp}}$, $\sigma_{\text{KLD}}$, $\sigma_{\text{aff}}$, and $\sigma_{2}$, in terms of the tightness of the lower bound. 
In the case of a complex model, where the formulation of the optimization problem might not be practical due to the number of constraints, or in case the full topology is not known, we show numerically that assuming a simpler underlying topology can provide a lower bound on the total EPR.

The paper is organized as follows. In section~\ref{sec:model}, we describe our model system and the coarse-graining approach. The results are presented in section~\ref{sec:results}: We discuss the estimator in subsection~\ref{sec:bounding}, apply it to different systems in subsection~\ref{sec:examples}, demonstrate how the accuracy of the measured statistics affects the results of our estimator in subsection~\ref{sec:accuracy}, and finally, we show the results of the optimization problem assuming a simpler underlying model in subsection~\ref{sec:simple}. We conclude our findings in section~\ref{sec:conclusion}.

\section{\label{sec:model} Model}
We assume a continuous time Markov chain over a finite and discrete set of states $i=\{1,2,\dots,N\}$. A trajectory is described by a sequence of states and their corresponding residence times before a transition to the next state occurs.
Being a Markovian process, the jump probabilities depend only on the current state.  

The transition rates $w_{ij}$ from state $i$ to $j$ determine the time evolution of the probabilities for the system to be in each state, according to the Master equation $\frac{d}{dt}{\bm p}(t)^T={\bm p}(t)^T{\bm W}$, where $T$ is the transpose operator, and ${\bm W}$ is the rate matrix
\begin{equation}
    [{\bm W}]_{ij}=
    \begin{cases}
        w_{ij} & j\neq i\\
        -\lambda_i & j=i
    \end{cases}
\end{equation}
${\bm p}(t)$ is a column vector of the state probabilities at time $t$, with $\sum_{i}p_i(t)=1$, and the diagonal entries are calculated according to $\lambda_i=\sum_{j\neq i}w_{ij}$ for probability conservation.

At the long-time limit, the system eventually reaches a steady state ${\bm \pi}$, where $\lim_{t\to\infty}p_i(t)=\pi_i$
such that $0=\frac{d}{dt}{\bm \pi}^T={\bm \pi}^T{\bm W}$ \cite{schnakenberg1976network}.

The waiting time 
at each state $i$ is an exponential random variable with mean waiting time of $\tau_i=\lambda_i^{-1}$. 

The mass rates $n_{ij}$ are defined as follows:
\begin{equation}
    n_{ij}=
    \begin{cases}
        \pi_iw_{ij} & j\neq i\\
        0 & j=i
    \end{cases}
    \label{eq:mass rates}
\end{equation}
The probabilities of jumping from state $i$ to state $j$ can be written in terms of the mass transition rates:
\begin{equation}
    p_{ij}=\dfrac{w_{ij}}{\lambda_i}=\dfrac{n_{ij}}{\sum_{j'\neq i}n_{ij'}}
    \label{eq:jump probabilities}
\end{equation}

The steady-state total EPR can be calculated by multiplying the net currents and the mass rate ratios (affinities), summing over all the links\cite{seifert2012stochastic, van2015ensemble}:
\begin{equation}
    \begin{split}
        \sigma_{\text{tot}} & =\sum_{i,j}\pi_iw_{ij}\log\left(\dfrac{\pi_iw_{ij}}{\pi_jw_{ji}}\right) \\
        & =\sum_{i,j}n_{ij}\log\left(\dfrac{n_{ij}}{n_{ji}}\right) \\
        & =\sum_{i<j}(n_{ij}-n_{ji})\log\left(\dfrac{n_{ij}}{n_{ji}}\right)
    \end{split}
\end{equation}
Given a long trajectory of a total duration $T$, the steady-state probability $\pi_i$ is the fraction of time spent in state $i$, and the mass rate $n_{ij}$ is the number of transitions $i\to j$ divided by $T$.

according to the definition of the mass transition rates in Eq.~\ref{eq:mass rates}, at the steady state, a mass conservation is satisfied at each state:
\begin{equation}
    \forall_i: \sum_jn_{ij}=\sum_jn_{ji}
    \label{eq:mass conservation}
\end{equation}

\begin{figure*}
\includegraphics{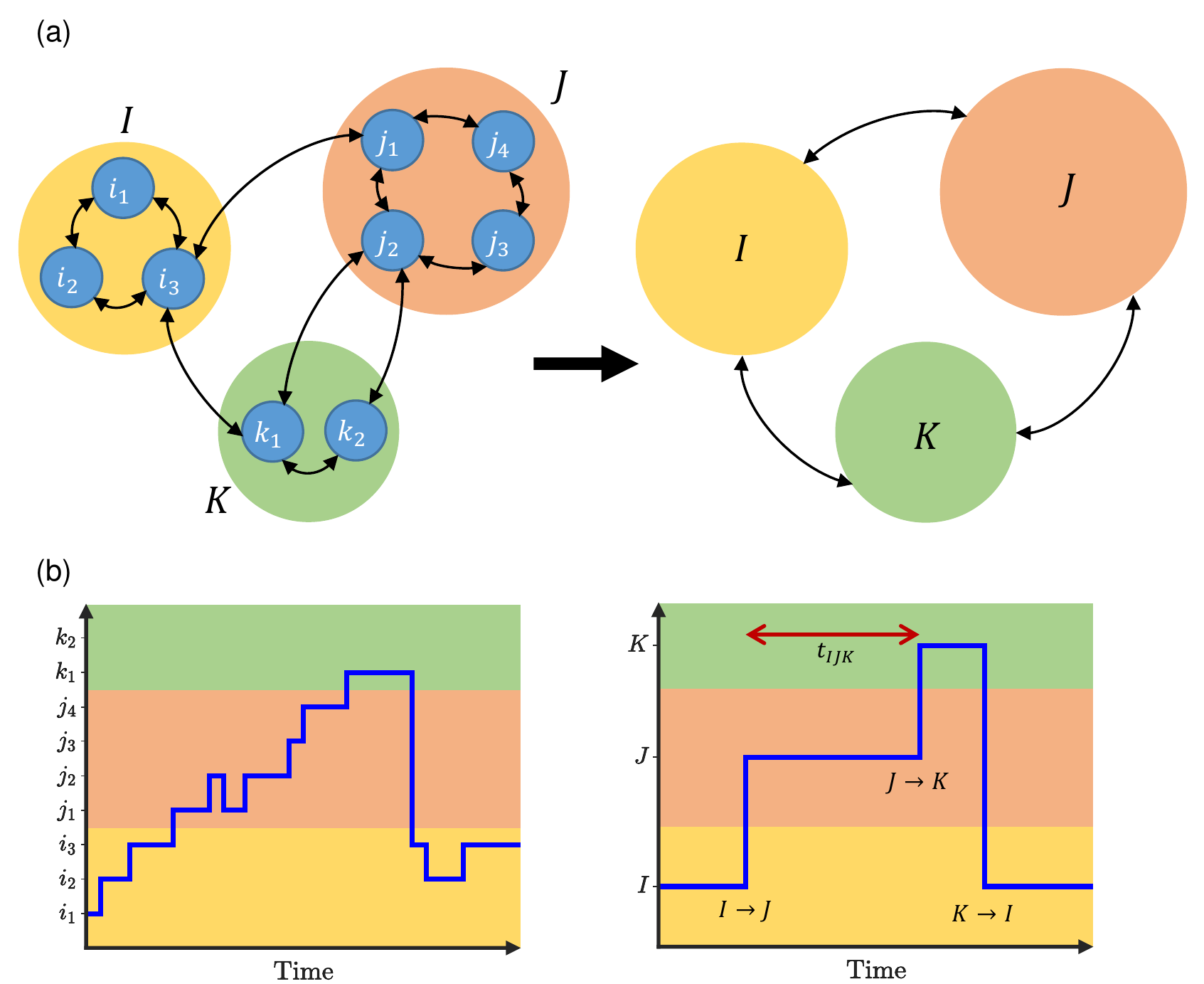}
\caption{\label{fig:fig1} Coarse graining. (a) The full Markovian system (left) and the coarse-grained system (right). (b) An example for a full trajectory (left) containing the actual states and the corresponding coarse-grained trajectory (right) containing only the observed states.}
\end{figure*}

In many practical scenarios, some of the microstates cannot be distinguished, and the transitions between them cannot be observed.
In such a case, a set of states $\{i_1,i_2,\dots,i_{N_I}\}$ is observed as a single coarse-grained state $I$ (Fig.~\ref{fig:fig1}(a)). The observed trajectory, therefore, includes only coarse-grained states and the combined residence time (Fig.~\ref{fig:fig1}(b)), and it is not necessarily a Markovian process \cite{martinez2019inferring}. Such a decimation procedure of lumping several states can give rise to semi-Markovian processes of any order depending on the topology of the network \cite{maes2009dynamical, zhang2019markovian, skinner2021improved, hartich2021violation}. In this case, the observed statistics of two or more consecutive transitions may give us additional information on the process.

\section{\label{sec:results} Results}
\subsection{\label{sec:bounding} Bounding the entropy production rate}
Given a coarse-grained system with a model of the full underlying Markovian network topology, we can formulate an optimization problem for obtaining a tight bound on the total EPR.
We consider a few observables: the coarse-grained steady-state probabilities, $\pi_I$, which is the probability to observe the system in the coarse-grained state $I$; the first-order mass transition rates, $n_{IJ}$, which is the rate of observing the transition $I\to J$; the second order mass transition rates, $n_{IJK}$, which is the rate of observing the transition $I\to J$ followed by the transition $J\to K$; and the conditional waiting time distributions $\psi_{IJK} (t)$, which is the distribution of waiting times in a coarse-grained state $J$ before a transition to a coarse-grained state $K$ occurs, conditioned on the previous transition being $I\to J$. 

We search over the space of all possible underlying systems with the same topology as our hypothesized Markovian model that give rise to the same observed statistics, while minimizing the EPR. Trivially, the EPR of the coarse-grained system at hand is bounded from below by the EPR of the underlying Markovian system with the same observed statistics after coarse-graining, having the minimal value of entropy production.

\subsubsection{Analytical expressions of the observed statistics}
The observed statistics of the coarse-grained system can be expressed analytically in terms of the mass rates and steady-state probabilities of the model underlying system. From probability and mass conservation, $\pi_I=\sum_{i\in I}\pi_i$, and $n_{IJ}=\sum_{i\in I,j\in J}n_{ij}$, respectively.
The mass conservation for the second-order transitions $n_{IJK}$ must include all the paths starting at state $i\in I$, passing through a state in $J$, where any number of transitions might occur inside $J$, and jumping to state $k\in K$.
To account for the transitions within $J$, we define the matrix ${\bm P}_{JJ}$ of the transition probabilities between states in $J$, $j_m,j_n\in J$:
\begin{equation}
    [{\bm P}_{JJ}]_{mn}=
    \begin{cases}
        p_{j_mj_n} & m\neq n\\
        0 & m=n
    \end{cases}
\end{equation}
Summing over the possible transitions from  $I$, transitions within $J$, and transitions to  $K$, we have (see Appendix~\ref{app:appendixA}):
\begin{equation}
    n_{IJK}=\sum_{i\in I, k\in K}{\bm n}_{iJ}^{T}[\mathbb{I}-{\bm P}_{JJ}]^{-1}{\bm p}_{Jk}
\end{equation}
where $\mathbb{I}$ is the identity matrix of the size of ${\bm P}_{JJ}$, and ${\bm n}_{iJ}$ and ${\bm p}_{Jk}$ are column vectors of the mass transition rates from state $i\in I$ to any state $j\in J$, and jump probabilities from any state $j\in J$ to a state $k\in K$, respectively:
\begin{equation}
    {\bm n}_{iJ}^T=[n_{ij_1}, n_{ij_2}, \cdots, n_{ij_{N_J}}]
\end{equation}
and:
\begin{equation}
    {\bm p}_{Jk}^T=[p_{j_1k}, p_{j_2k}, \cdots, p_{j_{N_J}k}]
\end{equation}

The conditional waiting time distribution $\psi_{IJK}(t)$ can be calculated by the Laplace and inverse-Laplace transforms (full derivations can be found in Appendix~\ref{app:appendixB}). We start from the Laplace transform of $\psi_{ij}(t)=w_{ij}e^{-\lambda_it}$, the joint probability distribution of the transition $i\to j$ and the waiting time in the Markovian state $i$:
\begin{equation}
    \Tilde{\psi}_{ij}(s)=\mathcal{L}\{\psi_{ij}(t)\}=\int_0^\infty\psi_{ij}(t)e^{-ts}dt=\dfrac{w_{ij}}{s+\lambda_i}
\end{equation}
Note that for any function $f(t)$, $\Tilde{f}(s\to 0)=\int_0^{\infty}f(t)e^{-ts}dt|_{s\to 0}=\int_0^{\infty}f(t)dt$ is the normalization of $f(t)$. Here, $\psi_{ij}(t)$ is normalized to $p_{ij}$, \emph{i.e.}, $p_{ij}=\int_0^\infty\psi_{ij}(t)dt$ (Eq.~\ref{eq:jump probabilities}).

Now, we consider the simple case where the second-order transition through the coarse-grained state $J$ starts and ends in specific Markovian states $i\in I$ and $k\in K$, respectively. The Laplace transform of the distribution of waiting times in $J$ before a transition to $k$ occur, given the previous transition was $i\to J$ is:
\begin{equation}
    \Tilde{\psi}_{iJk}(s)=\dfrac{{\bm p}_{iJ}^T}{\sum_{j\in J}p_{ij}}[\mathbb{I}-\Tilde{{\bm \Psi}}_{JJ}(s)]^{-1}\Tilde{{\bm \psi}}_{Jk}(s)
\end{equation}
where
\begin{equation}
    \Tilde{{\bm \psi}}_{Jk}^T(s)=[\Tilde{\psi}_{j_1k}(s), \Tilde{\psi}_{j_2k}(s), \cdots, \Tilde{\psi}_{j_{N_j}k}(s)]
    \label{eq:laplace vector}
\end{equation}
and $\Tilde{{\bm \Psi}}_{JJ}(s)$ is a matrix of the Laplace transforms of every joint probability distribution of waiting times and transitions within $J$:
\begin{equation}
    \Tilde{{\bm \Psi}}_{JJ}(s)=
    \begin{cases}
        \Tilde{\psi}_{j_mj_n}(s) & m\neq n\\
        0 & m=n
    \end{cases}
\end{equation}

We denote $\Tilde{\psi}_{iJK}(s)\equiv\sum_{k\in K}\Tilde{\psi}_{iJk}(s)$. Then, the Laplace transform of the conditional waiting time distribution is:
\begin{equation}
    \Tilde{\psi}_{IJK}(s)=\sum_{i\in I}\dfrac{\pi_i}{\pi_I}\dfrac{\Tilde{\psi}_{iJK}(s)}{\Tilde{\psi}_{iJK}(s\to 0)}
\end{equation}

Finally, we apply an inverse Laplace transform to obtain the conditional probability density:
\begin{equation}
    \psi_{IJK}(t)=\mathcal{L}^{-1}\{\psi_{IJK}(s)\}
\end{equation}

We further impose mass conservation at each of the Markovian states according to Eq.~\ref{eq:mass conservation}, to make sure the solution represents a valid Markovian system.

\subsubsection{Formalizing the optimization problem}
Let ${\mathcal S}$ be the real underlying Markovian system and let ${\mathcal R}$ be a general underlying system with the same topology as ${\mathcal S}$, i.e., the same states and possible transitions as ${\mathcal S}$, but ${\mathcal R}$ can have arbitrary mass rates and steady-state probabilities. 
Given the set of all systems ${\mathcal R}$ with the same steady-state probabilities $\pi_I^{\mathcal R}=\pi_I^{\mathcal S}$, same first-order mass transition rates $n_{IJ}^{\mathcal R}=n_{IJ}^{\mathcal S}$, same second-order mass transition rates $n_{IJK}^{\mathcal R}=n_{IJK}^{\mathcal S}$, and the same conditional waiting time distributions $\psi_{IJK}^{\mathcal R}(t)=\psi_{IJK}^{\mathcal S}(t)$, as the system ${\mathcal S}$, the following inequality holds for the EPR of ${\mathcal S}$ and ${\mathcal R}$, $\sigma({\mathcal S})$ and $\sigma({\mathcal R})$, respectively:
\begin{equation}
    \begin{split}
        \sigma_{\text{tot}}({\mathcal S})\geq\min_{\mathcal R}\{\sigma_{\text{tot}}({\mathcal R})|&\forall_{I,J,K}:\pi_I^{\mathcal R}=\pi_I^{\mathcal S}, n_{IJ}^{\mathcal R}=n_{IJ}^{\mathcal S}, \\
        &n_{IJK}^{\mathcal R}=n_{IJK}^{\mathcal S}, \\
        &\psi_{IJK}^{\mathcal R}(t)=\psi_{IJK}^{\mathcal S}(t)\}\equiv\sigma_{\text{opt}}^{(\infty)}
    \end{split}
\end{equation}
where $\sigma_{\text{opt}}^{(\infty)}$ is the minimal EPR value of all the possible underlying systems ${\mathcal R}$. The inequality holds since the real system ${\mathcal S}$ belongs to the set of systems over which we minimize.
The only variables of the optimization problem are $n_{ij}$ and $\pi_i$, from which one can fully describe any of the possible underlying Markovian systems ${\mathcal R}$.
All the constraints, $\pi_I$, $n_{IJ}$, $n_{IJK}$, and $\psi_{IJK}(t)$, as well as the EPR objective function, depend on these variables.
Note that these variables are bounded by $0\leq \pi_i \leq \pi_I$ and $0\leq n_{ij} \leq n_{IJ}$.

In contrast to the constraints on the steady-state probabilities and the first- and second-order mass transition rate values, the constraint on the waiting-time distributions requires an equality of continuous functions $\psi_{IJK}(t)$, which one cannot fully reconstruct from trajectory data of finite duration. Moreover, solving the optimization problem using a constraint on a function with non-trivial dependency on the optimization problem variables is extremely challenging. Thus, we modify the optimization, and instead, use the moments of the waiting time distributions:
\begin{equation}
    \begin{split}
        \sigma_{\text{opt}}^{(n)}({\mathcal S})\equiv\min_{\mathcal R}\{\sigma_{\text{tot}}({\mathcal R})|&\forall_{I,J,K}:\pi_I^{\mathcal R}=\pi_I^{\mathcal S}, n_{IJ}^{\mathcal R}=n_{IJ}^{\mathcal S}, \\
        &n_{IJK}^{\mathcal R}=n_{IJK}^{\mathcal S}, \\
        &\forall_{k\in \{1,2,...,n\}}:\langle t_{IJK}^k\rangle^{\mathcal R}=\langle t_{IJK}^k\rangle^{\mathcal S}\}
    \end{split}
\end{equation}
where $\langle t_{IJK}^k\rangle$ is the $k$-th moment of the conditional waiting time distribution $\psi_{IJK}(t)$. Using increasing number of moments, we can write the hierarchical bounds:
\begin{equation}
    \forall_{n\in \mathbb{N}}:\sigma_{\text{tot}}({\mathcal S})\geq \sigma_{\text{opt}}^{(\infty)}({\mathcal S})\geq \sigma_{\text{opt}}^{(n)}({\mathcal S})\geq\cdots\geq \sigma_{\text{opt}}^{(1)}({\mathcal S})
\end{equation}

We can easily get the analytical expressions for the moments $\langle t_{IJK}^k\rangle$ from the Laplace transform {(see Appendix~\ref{app:appendixB})}:
\begin{equation}
    \langle t_{IJK}^k\rangle=(-1)^k\dfrac{d^k\Tilde{\psi}_{IJK}(s)}{ds^k}|_{s\to 0}
\end{equation}

Now, for each moment, we have an expression that depends on the optimization problem variables in a simpler way, which in turn, simplifies the calculations.
After calculating the values of the observables for the optimization problem, we solve it using a global search non-linear optimization algorithm \cite{ugray2007scatter}.

\subsection{\label{sec:examples} Examples}
\subsubsection{\label{sec:4state} 4-state system}
\begin{figure}
\includegraphics{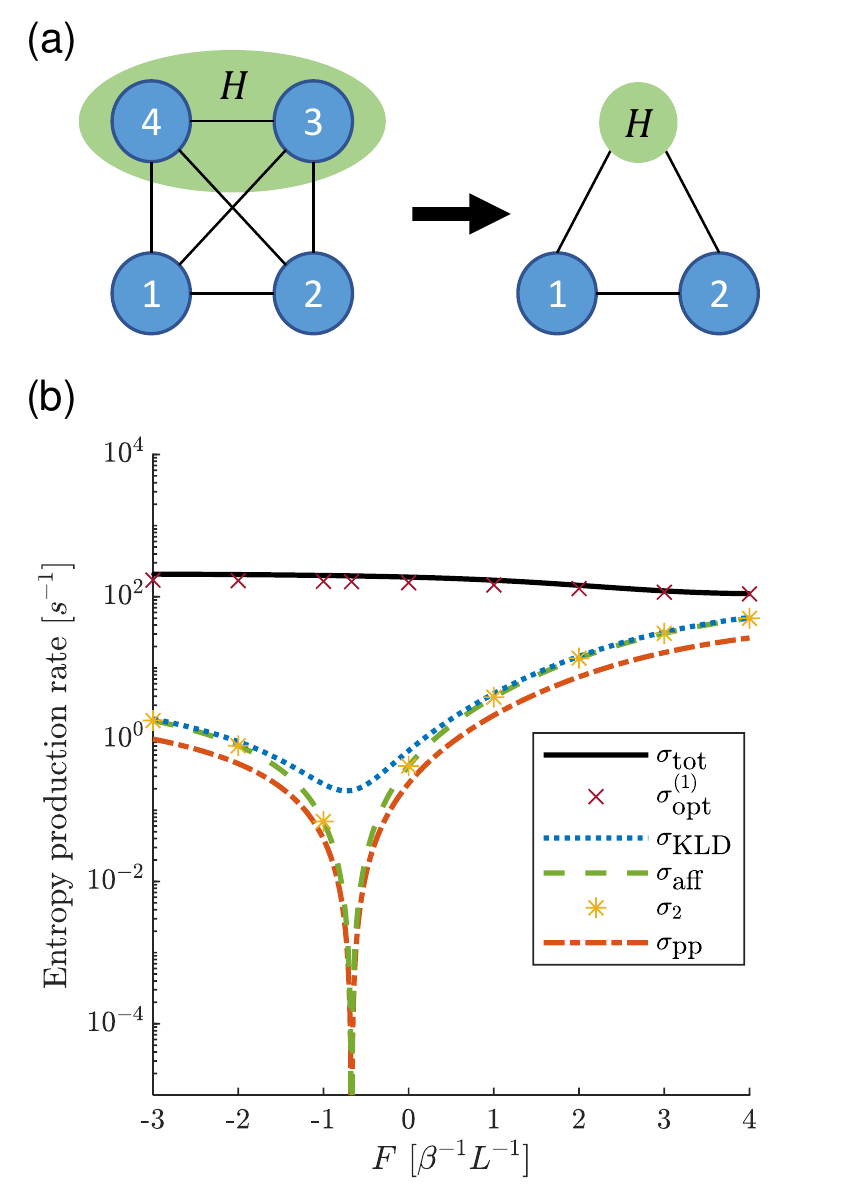}
\caption{\label{fig:fig2} 4-state system. (a) Illustration of the full 4-state system topology, including the coarse-graining of states 3 and 4 to state $H$. (b) Total EPR $\sigma_{\text{tot}}$ (solid black line), our bound $\sigma_{\text{opt}}^{(1)}$ (brown cross), KLD estimator $\sigma_{\text{KLD}}$ (dotted blue line), affinity estimator $\sigma_{\text{aff}}$ (dashed green line), two-step estimator $\sigma_2$ (yellow Asterisk), and the passive partial entropy production $\sigma_{\text{pp}}$ (dashed-dotted orange line). The rates we used are $w_{12}=3\ s^{-1}$, $w_{13}=0\ s^{-1}$, $w_{14}=8\ s^{-1}$, $w_{21}=2\ s^{-1}$, $w_{23}=50\ s^{-1}$, $w_{24}=0.2\ s^{-1}$, $w_{31}=0\ s^{-1}$, $w_{32}=2\ s^{-1}$, $w_{34}=75\ s^{-1}$, $w_{41}=1\ s^{-1}$, $w_{42}=35\ s^{-1}$, $w_{43}=0.7\ s^{-1}$.}
\end{figure}

We consider a fully-connected network of 4 states, with two observed states $\{1, 2\}$ and two hidden states $\{3, 4\}$, which are coarse-grained to state $H$ (Fig.~\ref{fig:fig2}(a)), resulting in second-order semi-Markov dynamics \cite{martinez2019inferring}. 
The observed statistics of interest are the steady state probabilities $\pi_1, \pi_2$ and $\pi_H$, the first-order mass transition rates $n_{1H}$, $n_{H1}$, $n_{2H}$, and $n_{H2}$, the second-order mass transition rates $n_{1H2}$ and $n_{2H1}$ and the ${k}$-th moment of the conditional waiting time distributions $\langle t_{1H1}^k\rangle$, $\langle t_{1H2}^k\rangle$, $\langle t_{2H1}^k\rangle$ and $\langle t_{2H2}^k\rangle$. Notice we only used the second-order statistics through the coarse-grained state $H$, since states $1$ and $2$ are Markovian. Furthermore, we do not use $n_{1H1}$ and $n_{2H2}$ since they depend on the other mass transition rates: $n_{1H1}=n_{1H}-n_{1H2}$ and $n_{2H2}=n_{2H}-n_{2H1}$.
The derivations of the analytical expressions of the second-order mass transition rates and the moments of the conditional waiting time moments, for this system, can be found in Appendix~\ref{app:appendixC}.

We tune the transition rates over the observed link between states $1$ and $2$ according to $w_{12}(F)=w_{12}e^{-\beta FL}$ and $w_{21}(F)=w_{21}e^{\beta FL}$, where $\beta=T^{-1}$ is the inverse temperature (with $k_{\text B}=1$), and $L$ is a characteristic length scale, to mimic external forcing. We compare the different EPR estimators on the system for several values for a driving force $F$ over the observed link (Fig.~\ref{fig:fig2}(b)).

The passive partial EPR \cite{bisker2017hierarchical}:
\begin{equation}
    \begin{split}
        \sigma_{\text{pp}}&=\left(\pi_1 w_{12} - \pi_2 w_{21}\right)\log\left(\dfrac{\pi_1 w_{12}}{\pi_2 w_{21}}\right) \\
        &=(n_{12}-n_{21})\log\left(\dfrac{n_{12}}{n_{21}}\right)
    \end{split}
\end{equation}
The KLD estimator is the sum of two contributions:
\begin{equation}
    \begin{split}
        \sigma_{\text{KLD}}&=\sigma_{\text{aff}}+\sigma_{\text{WTD}} \\
        &
        \begin{split}
            &=\dfrac{1}{\mathcal{T}}\sum_{I,J,K}p_{IJK}\log\left(\dfrac{p([IJ]\to [JK])}{p([KJ]\to [JI])}\right) \\
            &+\dfrac{1}{\mathcal{T}}\sum_{I,J,K}p_{IJK}D\left[\psi_{IJK}(t)||\psi_{KJI}(t)\right]
        \end{split}
    \end{split}
\end{equation}
where $p([IJ]\to [JK])$ is the probability to observe the transition $J\to K$ given the previous transition was $I\to J$, $p_{IJK}$ is the probability to observe the second-order transition $I\to J\to K$, and $D[p||q]$ is the KLD between the probability distributions $p$ and $q$. As was previously shown, the hierarchy between the EPR estimators is $\sigma_{\text{KLD}}\geq\sigma_{\text{aff}}\geq\sigma_{\text{pp}}$ \cite{bisker2017hierarchical, martinez2019inferring}.

The $\sigma_{2}$ estimator is also formulated as an optimization problem searching over a canonical form of the system with the same observed statistics, however, it only considers the first- and second-order mass transition rates \cite{skinner2021improved}. Its place in the hierarchy between the EPR estimators varies for different systems. While $\sigma_{2}$ can be greater than $\sigma_{\text{KLD}}$ in some cases \cite{skinner2021improved}, here, for the rate values we used, $\sigma_{2}<\sigma_{\text{KLD}}$. In fact, although the values of $\sigma_{2}$ and $\sigma_{\text{aff}}$ appear to be similar (Fig.~\ref{fig:fig2}(b)), actually $\sigma_{2}<\sigma_{\text{aff}}$ for all of the values of $F$ used.

At the stalling force, there is no current in the visible link and we get $\sigma_{\text{pp}}=\sigma_{\text{aff}}=\sigma_{2}=0$, which is the trivial bound.
In contrast, $\sigma_{\text{KLD}}$ and our estimator $\sigma_{\text{opt}}^{(1)}$ give a non-trivial bound. Moreover, $\sigma_{\text{opt}}^{(1)}$ surpasses $\sigma_{\text{KLD}}$ significantly and yields a tight bound. For this system, using higher moments in order to calculate $\sigma_{\text{opt}}^{(2)}$ did not make any improvement compared to $\sigma_{\text{opt}}^{(1)}$.

\begin{figure*}
\includegraphics{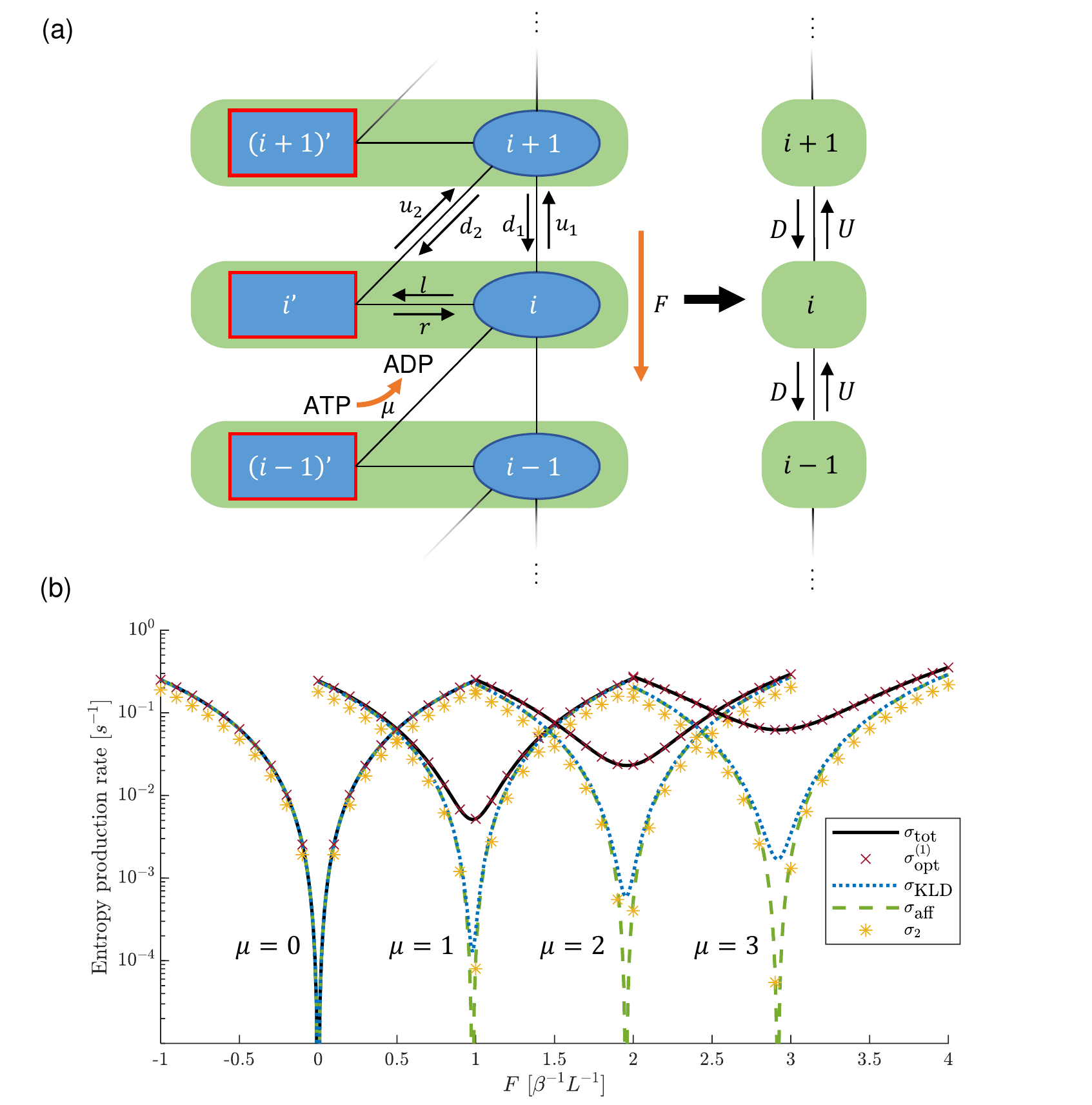}
\caption{\label{fig:fig3} Molecular motor. (a) Illustration of the full molecular motor system including the coarse-graining of the active (red boxed square) and passive (ellipse) states. (b) Total EPR $\sigma_{\text{tot}}$ (solid black line), our bound $\sigma_{\text{opt}}^{(1)}$ (brown cross), KLD estimator $\sigma_{\text{KLD}}$ (dotted blue line), the affinity estimator $\sigma_{\text{aff}}$ (dashed green line), and the two-step estimator $\sigma_2$ (yellow Asterisk). The rates we used are $w_r=w_l=w_{u2}=w_{d2}=1\ s^{-1}$, $w_{u1}=w_{d1}=0.01\ s^{-1}$.}
\end{figure*}

\begin{figure*}
\includegraphics{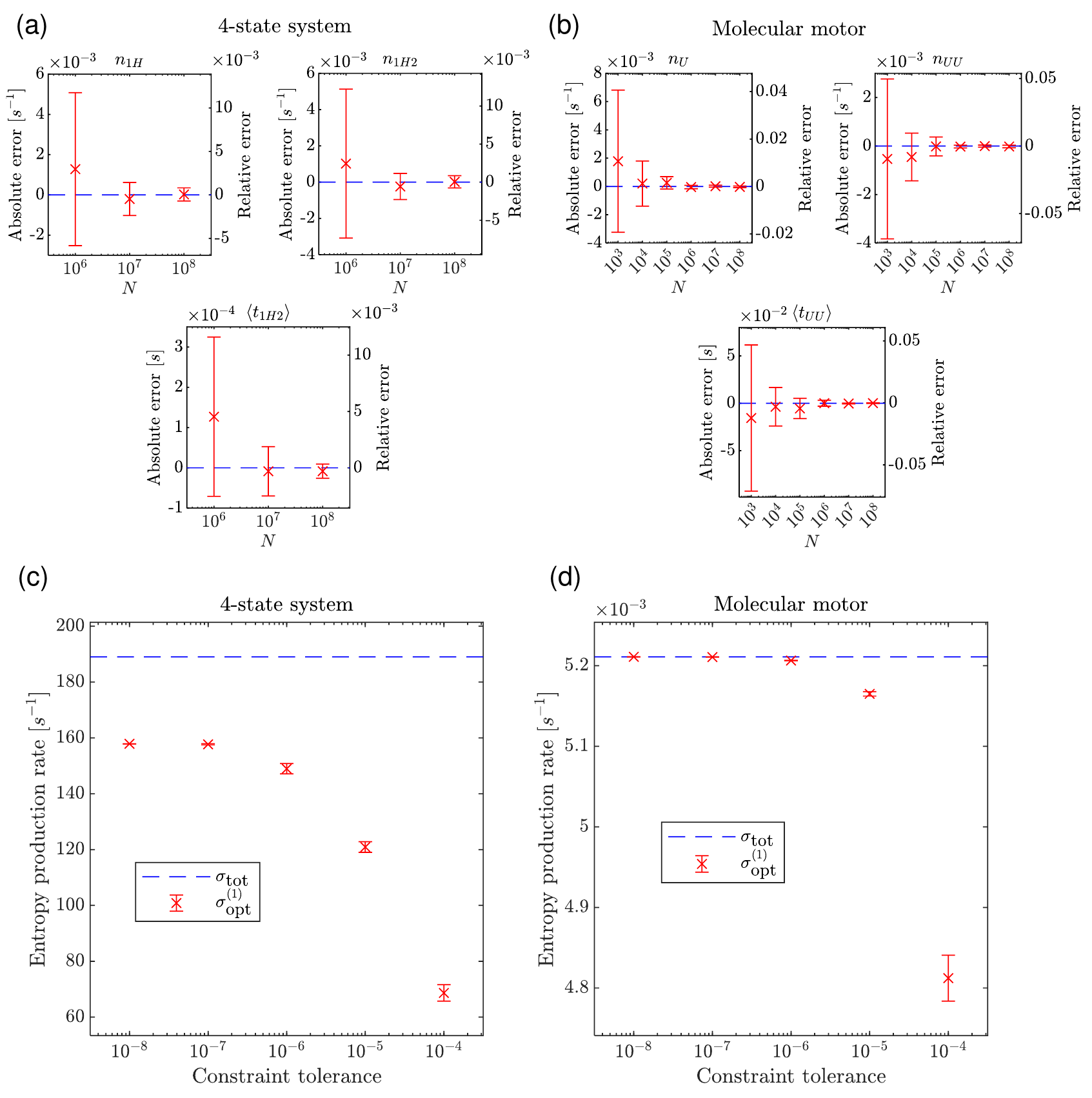}
\caption{\label{fig:fig4} Importance of data accuracy. (a) The error of some statistics of the 4-state system for different values of the trajectory length $N$. The absolute and relative errors are on the left and right axes, respectively. 
(b) The error of some statistics of the molecular motor system for different values of the trajectory length $N$. The absolute and relative errors are on the left and right axes, respectively. 
(c) The error of $\sigma_{\text{opt}}^{(1)}$ results for the 4-state system for different constraint tolerance values, using the analytical statistics values. 
(d) The error of $\sigma_{\text{opt}}^{(1)}$ results for the molecular motor system for different constraint tolerance values, using the analytical statistics values. 
Error bars stand for the standard deviation of 10 different realizations.}
\end{figure*}

\subsubsection{\label{sec:motor} Molecular motor}
Here, we study a model of a molecular motor, illustrated in Fig.~\ref{fig:fig3}(a). The motor can physically move in space (upward or downward), $i\leftrightarrow i+1$, or change internal states (passive or active), $i\leftrightarrow i'$. An external source of chemical work $\Delta\mu$ drives the upward spatial jumps from the active state, and a mechanical force $F$ acts against it and drives the downward transitions. We assume that an external observer cannot distinguish between the internal states of the motor, but rather can only record its physical position. The observed statistics are thus of a second-order Semi-Markov process \cite{martinez2019inferring}.

Owing to the transnational symmetry in the model, we represent the molecule motor as a cyclic network of three coarse-grained states where each of them represents the physical location, lumping the active and passive internal states.  We denote the steady-state probability of being in the passive and active states as $\pi$ and $\pi'$, respectively. Notice that the probability to be in each physical location in the $3$-state cyclic system is the same, and that $\pi$ and $\pi'$ are the same for all of the physical locations, therefore, $\pi+\pi'=1/3$.

We denote the upward and downward transitions from and to the passive state as $u_1$ and $d_1$, respectively, the upward and downward transitions from and to the active state as $u_2$ and $d_2$, respectively, and the transitions between the active and passive states at the same physical location as $r$ (right) and $l$ (left), respectively. The upward and downward coarse-grained transitions are labeled as $U$ and $D$, respectively.

The observed statistics of interest are the first-order mass rates $n_U$, $n_D$, the second-order mass rates $n_{UU}$, $n_{DD}$ and the $k$-th moment of the conditional waiting times $\langle t_{UU}^k\rangle$, $\langle t_{UD}^k\rangle$, $\langle t_{DU}^k\rangle$ and $\langle t_{DD}^k\rangle$. 
Note that we do not use $n_{UD}$ and $n_{DU}$, since they depend on the other mass rates: $n_{UD}=n_{U}-n_{UU}$ and $n_{DU}=n_{D}-n_{DD}$. 
Owing to the symmetry of the cycle representation of the coarse-grained system, in which the steady-state probabilities are equally distributed, we only need the constraints on the upward and downward transitions.
The derivations of the analytical expressions of the second-order mass transition rates and the moments of the conditional waiting time distributions, for this system, can be found in Appendix~\ref{app:appendixD}.

The chemical affinity $\mu$, arising from ATP hydrolysis for example, only affects the transitions $u_2$ and $d_2$, whereas the external force $F$ affects all of the spatial transitions $u_1$, $d_1$, $u_2$ and $d_2$. The transition rates then obey local detailed balance: $w_{d1}/w_{u1}=e^{\beta FL}$ and $w_{d2}/w_{u2}=e^{\beta(FL-\mu)}$, where $L$ is the length of a single spatial jump \cite{martinez2019inferring}.

We compare the different EPR estimators for the molecular motor system for several values of $\mu$ and for each $\mu$ value, we tune the external forcing parameter $F$ (Fig.~\ref{fig:fig3}(b)). Notice the passive partial EPR, $\sigma_{\text{pp}}$, is not applicable for this system since all  the original Markovian states are coarse-grained.

The hierarchy of the different EPR estimators for the molecular motor, for the rate values we used, is $\sigma_{\text{opt}}^{(1)}\geq \sigma_{\text{KLD}} \geq \sigma_{\text{aff}} \geq \sigma_{2}$. At the stalling force for each value of $\mu$, where there is no visible current, we find $\sigma_{\text{aff}}=\sigma_{2}=0$,  which is the trivial bound. In contrast, similar to the 4-state system, $\sigma_{\text{opt}}^{(1)}$ surpasses $\sigma_{\text{KLD}}$ significantly and yields a tight bound.

\begin{figure*}
\includegraphics{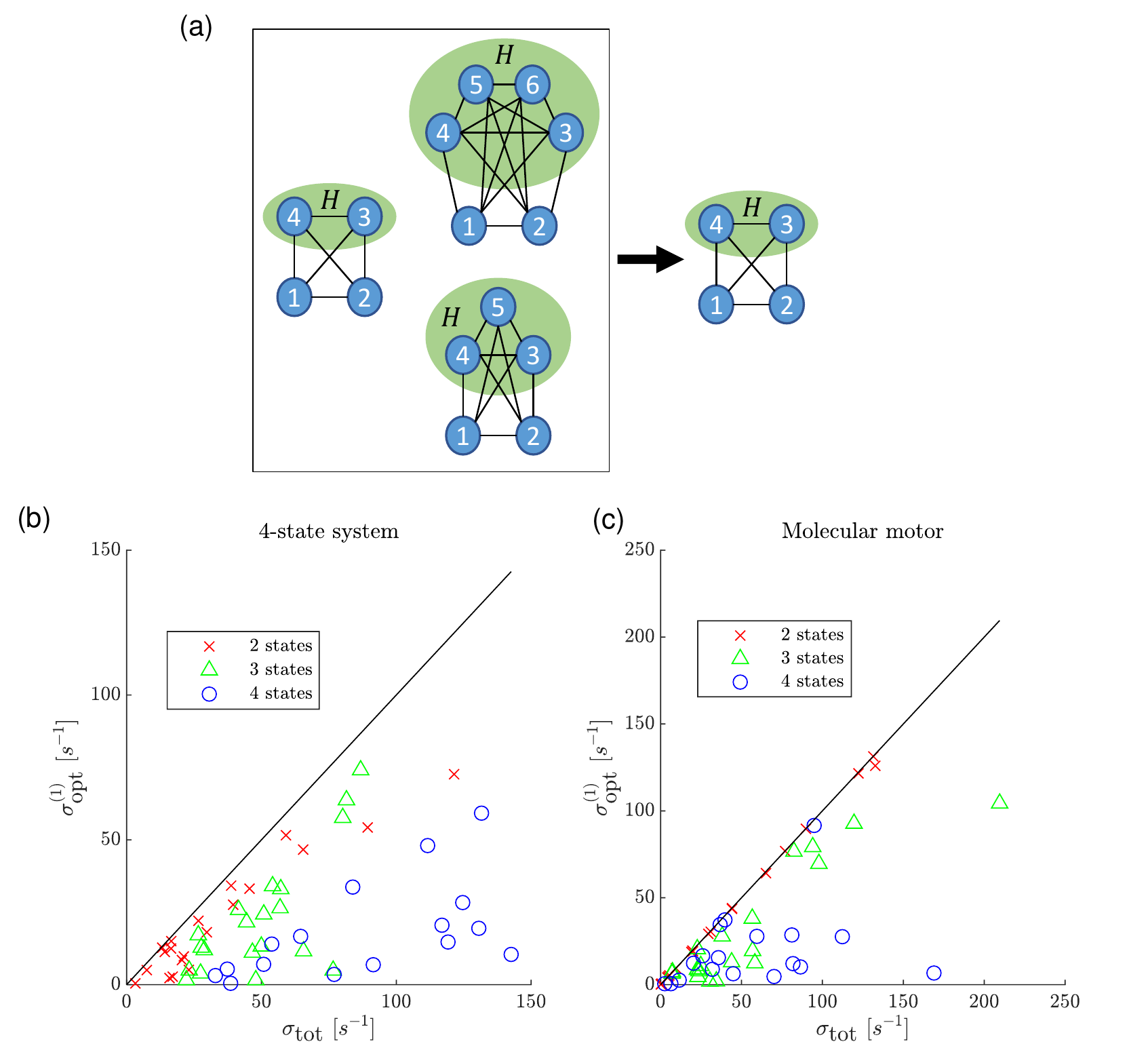}
\caption{\label{fig:fig5} Optimizing using a simple model. (a) Illustration of solving the optimization problem for a simple model with 2 hidden states (right), whereas the real system has more hidden states (left). (b) The results of $\sigma_{\text{opt}}^{(1)}$ assuming the simple 4-state model (2 hidden states), when the real system has 2 (red cross), 3 (green triangle) or 4 (blue circle) hidden states. (c) The results of $\sigma_{\text{opt}}^{(1)}$ assuming the simple molecular motor model (2 hidden states), when the real system has 2 (red cross), 3 (green triangles) or 4 (blue circle) hidden states. For both systems, the results are presented for random generated transition rates (for each case) with statistics calculated from trajectories of length $N=10^8$ using a constraint tolerance of $10^{-5}$.}
\end{figure*}

\subsection{\label{sec:accuracy} Importance of data accuracy}
One of the hyper parameters defining the optimization problem is the constraint tolerance, which indicates the acceptable numerical error of the solution. If $\epsilon$ is the absolute error of the trajectory statistics with respect to the true analytical ones, then the constraint tolerance must be equal to or greater than $\epsilon$. Otherwise, the optimization problem might not converge or give an overestimate in the worst-case scenario.

In Fig.~\ref{fig:fig4}, we plot the absolute (and relative) error of a few statistics values calculated from several trajectories as function of the trajectory length $N$, for both systems discussed in the previous sections. Moreover, using the analytical values of the statistics for maximum accuracy, we plot the results of our estimator $\sigma_{\text{opt}}^{(1)}$ as function of the constraint tolerance.

As expected, longer trajectory data result in a more accurate estimation of the observed statistics used for our optimization problem for both systems, as evident from the values of $n_{1H}$, $n_{1H2}$ and $\langle t_{1H2} \rangle$ for the 4-state system (Fig.~\ref{fig:fig4}(a)), and from the values of $n_U$, $n_{UU}$, and $\langle t_{UU} \rangle$ for the molecular motor (Fig.~\ref{fig:fig4}(b)). For smaller errors, we can use a smaller constraint tolerance.

For both systems, smaller constraint tolerance leads to a better estimator as the value of the lower bound on the EPR approaches the true analytical value (Fig.~\ref{fig:fig4}(c) and (d)), demonstrating the importance of an accurate estimation of the observables.

\subsection{\label{sec:simple} Optimizing a simple model}
Although our approach can be generalized to any number of hidden states, the analytical expressions for the observables become complicated, and the number of variables increases for a more complex coarse-grained topology. In turn, solving the optimization problem would require longer computation times. In order to test the performance of our estimator, we solved the optimization problem for a larger number of hidden states in a fully-connected network of $4$, $5$, and $6$ states with only $2$ observed states, assuming only $2$ states are coarse-grained (Fig.~\ref{fig:fig5}(a)). Similarly, we tested the performance of our estimator for the case of the molecular motor with $2$, $3$, and $4$ internal states at each physical position, assuming there are only $2$. While generally, the estimator gives a more accurate result for the case of the $2$ hidden state, which matches the assumption, it still provides a lower bound on the total EPR with comparable accuracy for a larger number of hidden states in the two systems (Fig.~\ref{fig:fig5}(b) and (c)).

\section{\label{sec:conclusion} Conclusion}
We present a new estimator for the entropy production rate, which gives a tight bound by formulating an optimization problem using both transitions and waiting times statistics. Our estimator can be applied to any system with known topology and it significantly surpasses previous estimators, as demonstrated for the two studied systems, the fully-connected hidden network, and the molecular motor. The variables for the optimization problem can be inferred from the observed statistics, where longer trajectories result in more accurate estimation and enable a smaller constraint tolerance value.
Finally, for both systems, our approach can provide a lower bound on the total EPR for more complex systems, assuming a simpler underlying topology of the hidden states. Although we numerically showed that searching over all the systems with a simpler topology of the hidden part and the same observed statistics as the true system gave a lower bound on the total EPR for the two systems we studied, it remains an open problem to show this approach is universal. It would be interesting for future work to determine whether removing states from the hidden sub-network can only decrease the entropy production, given the observed statistics are conserved.

In summary, our approach is based on an optimization problem formulated using the observed statistics of a partially accessible system and provides a tight lower bound on the total EPR. The estimator can be used as a benchmark for comparing the performance of other estimators that rely on coarse-grained or partial information about the system.

\begin{acknowledgments}
G. Bisker acknowledges the Zuckerman STEM Leadership Program, and the Tel Aviv University Center for AI and Data Science (TAD). This work was supported by the ERC NanoNonEq 101039127, the Air Force Office of Scientific Research (AFOSR) under award number FA9550-20-1-0426, and by the Army Research Office (ARO) under Grant Number W911NF-21-1-0101. The views and conclusions contained in this document are those of the authors and should not be interpreted as representing the official policies, either expressed or implied, of the Army Research Office or the U.S. Government.
\end{acknowledgments}

\appendix

\section{\label{app:appendixA} Second-order mass rates}
In order to find the second-order mass transition rates for two consecutive transitions between coarse-grained states, $n_{IJK}$, we need to take into account every possible original state $i\in I$, every possible path within the coarse grained state $J$, and every possible transition from a state in $J$ to every possible final state $k\in K$. Let us start by considering a specific initial Markovian state $i\in I$ and a specific final Markovian state $k\in K$ and calculate the mass transition rate $n_{iJk}$:
\begin{equation}
    \begin{split}
        n_{iJk}&=\sum_{N=0}^{\infty}\sum_{j_0,\dots,j_N\in J}n_{ij_0}p_{j_0j_1}p_{j_1j_2}\cdots p_{j_{N-1}j_N}p_{j_N k} \\
        &=\sum_{N=0}^{\infty}\sum_{j',j''\in J}n_{ij'}\left[{\bm P}_{JJ}^N\right]_{j'j''} p_{j'' k} \\
        &=\sum_{j',j''\in J}n_{ij'}\left(\sum_{N=0}^{\infty}\left[{\bm P}_{JJ}^N\right]_{j'j''}\right) p_{j'' k} \\
        &=\sum_{j',j''\in J}n_{ij'}\left[\mathbb{I}-{\bm P}_{JJ}\right]_{j'j''}^{-1} p_{j'' k} \\
        &={\bm n}_{iJ}^T\left[\mathbb{I}-{\bm P}_{JJ}\right]^{-1} {\bm p}_{J k}
    \end{split}
    \label{eq:second-order mass transition rates 1}
\end{equation}
The two summations are for all the possible lengths $N$ of trajectories within $J$, and all the optional paths with the given length $\{j_0,j_1,\cdots,j_N\}$ in $J$. From mass conservation, we can now obtain the expression for $n_{IJK}$ by summing over all the optional original $i\in I$ and final $k\in K$ states:
\begin{equation}
    n_{IJK}=\sum_{i\in I}\sum_{k\in K}n_{iJk}
    \label{eq:second-order mass transition rates 2}
\end{equation}

\section{\label{app:appendixB} Conditional waiting time moments}
The waiting time at each Markovian state $i$ is an exponentially distributed random variable $\psi_i(t)$ with mean waiting time $\tau_i=\lambda_i^{-1}$:
\begin{equation}
    \psi_i(t)=\lambda_i e^{-\lambda_it}
\end{equation}

For the calculations, we used the joint distribution of the waiting time and the transition $i\to j$:
\begin{equation}
    \psi_{ij}(t)=w_{ij}e^{-\lambda_it}
\end{equation}
Notice that $\psi_{ij}(t)$ is not normalized to 1 as $\int_0^\infty\psi_{ij}(t)dt=p_{ij}$. 

The probability to observe a trajectory $\gamma_N:i_0\to i_1\to\cdots\to i_N$ with a total duration of $T$ is:
\begin{equation}
    \begin{split}
        &p(\gamma_N,T)= \\
        &
        \begin{split}
            =\int_{\sum\limits_{i=0}^{N-1} t_i=T}&\psi_{i_0 i_1}(t_0)\psi_{i_1 i_2}(t_1)\cdots\psi_{i_{N-1} i_N}(t_{N-1}) \\ 
            &dt_0 dt_1\cdots dt_{N-1}
        \end{split}
    \end{split}
\end{equation}
Since this is a convolution, we can perform a Laplace transform to get a simpler formula of multiplications of Laplace transforms of Markovian joint distributions of waiting times and transitions: 
\begin{equation}
    \Tilde{p}(\gamma_N,s)=\Tilde{\psi}_{i_0 i_1}(s)\Tilde{\psi}_{i_1 i_2}(s)\cdots\Tilde{\psi}_{i_{N-1} i_N}(s)
\end{equation}
where
\begin{equation}
    \begin{split}
        \Tilde{\psi}_{ij}(s)&=\int_0^{\infty}\psi_{ij}(t)e^{-st}dt=\int_0^{\infty}w_{ij}e^{-\lambda_i t}e^{-st}dt \\
        &=w_{ij}\int_0^{\infty}e^{-(s+\lambda_i)t}dt=w_{ij}\left[-\dfrac{e^{-(s+\lambda_i)t}}{s+\lambda_i}\right]_0^{\infty} \\
        &=\dfrac{w_{ij}}{s+\lambda_i}
    \end{split}
    \label{eq:laplace markovian}
\end{equation}

In order to calculate the moments of the conditional waiting time distribution ${\psi}_{IJK}(t)$ for the coarse-grained state $J$ conditioned on an initial state in $I$ and a final state in $K$, our strategy is to calculate its Laplace transform $\Tilde{\psi}_{IJK}(s)$. We start by calculating $\Tilde{\psi}_{iJk}(s)$ which is the Laplace transform of the waiting distribution in coarse-grained state $J$, before jumping to a specific Markovian state $k\in K$, given it came from a specific Markovian state $i\in I$. 
Since we want the waiting time in $J$, we sum over all of the paths with any length $N$ inside $J$ with a final transition to $k\in K$, $j_0\to j_1\to\cdots\to j_N\to k$, weighed by the probability to jump from $i\in I$ to the first state $j_0\in J$:
\begin{equation}
    \begin{split}
        &\Tilde{\psi}_{iJk}(s)= \\
        &=
        \begin{split}
            \sum_{N=0}^{\infty}\sum_{j_0,\dots,j_N\in J}&\dfrac{p_{ij_0}}{\sum\limits_{j\in J}p_{ij}}\Tilde{p}(j_0\to j_1\to\cdots\to j_N\to k,s)
        \end{split} \\
        &=
        \begin{split}
            \sum_{N=0}^{\infty}\sum_{j_0,\dots,j_N\in J}&\dfrac{p_{ij_0}}{\sum\limits_{j\in J}p_{ij}}\Tilde{\psi}_{j_0 j_1}(s)\cdots\Tilde{\psi}_{j_{N-1} j_N}(s)\Tilde{\psi}_{j_N k}(s)
        \end{split} \\
        &=\sum_{N=0}^{\infty}\sum_{j',j''\in J}\dfrac{p_{ij'}}{\sum\limits_{j\in J}p_{ij}}\left[\Tilde{{\bm \Psi}}_{JJ}(s)^N\right]_{j',j''}\Tilde{\psi}_{j'' k}(s) \\
        &=\sum_{j',j''\in J}\dfrac{p_{ij'}}{\sum\limits_{j\in J}p_{ij}}\sum_{N=0}^{\infty}\left[\Tilde{{\bm \Psi}}_{JJ}(s)^N\right]_{j',j''}\Tilde{\psi}_{j'' k}(s) \\
        &=\sum_{j',j''\in J}\dfrac{p_{ij'}}{\sum\limits_{j\in J}p_{ij}}\left[\mathbb{I}-\Tilde{{\bm \Psi}}_{JJ}(s)\right]_{j',j''}^{-1}\Tilde{\psi}_{j'' k}(s) \\
        &=\dfrac{{\bm p}_{iJ}^T}{\sum\limits_{j\in J}p_{ij}}\left[\mathbb{I}-\Tilde{{\bm \Psi}}_{JJ}(s)\right]^{-1}\Tilde{{\bm \psi}}_{J k}(s)
    \end{split}
    \label{eq:laplace iJk}
\end{equation}
where $\Tilde{{\bm \Psi}}_{JJ}(s)$ is a matrix of size $N_J\times N_J$, and $N_J$ is the number of Markovian states inside $J$:
\begin{equation}
    \left[\Tilde{{\bm \Psi}}_{JJ}(s)\right]_{j_1,j_2}=
    \begin{cases}
        \Tilde{\Psi}_{j_1j_2}(s) & j_1\neq j_2\\
        0 & j_1=j_2
    \end{cases}
    \label{eq:laplace matrix}
\end{equation}

As mentioned in the main text we denote $\Tilde{\psi}_{iJK}(s)\equiv\sum_{k\in K}\Tilde{\psi}_{iJk}(s)$. Notice that $\Tilde{\psi}_{iJK}(s)$ is not normalized to 1 and it needs to be divided by $\Tilde{\psi}_{iJK}(s\to 0)$, which is exactly the probability to jump from $J$ to $K$, given the transition to $J$ was from $i$. 
\begin{equation}
    \Tilde{\psi}_{iJK}^{Normalized}(s)=\dfrac{\Tilde{\psi}_{iJK}(s)}{\Tilde{\psi}_{iJK}(s\to 0)}
\end{equation}
This results from the fact that we used $\psi_{ij}(t)$, which is normalized to $p_{ij}$.

In order to get $\Tilde{\psi}_{IJK}(s)$, we sum $\Tilde{\psi}_{iJK}^{Normalized}(s)$ over all of the Markovian states $i\in I$, weighed by the corresponding probability $\pi_i/\pi_I$ of being in state $i$, given the system is in the coarse-grained state $I$:
\begin{equation}
    \Tilde{\psi}_{IJK}(s)=\sum_{i\in I}\dfrac{\pi_i}{\pi_I}\Tilde{\psi}_{iJK}^{Normalized}(s)
    \label{eq:laplace IJK}
\end{equation}

For a general probability density function $f(t):[0,\infty]\to [0,1]$ the Laplace transform is:
\begin{equation}
    \Tilde{f}(s)=\int_0^{\infty}f(t)e^{-st}dt
\end{equation}
and its $k$-th derivative by $s$ is:
\begin{equation}
    \dfrac{d^k \Tilde{f}(s)}{ds^k}=(-1)^k\int_0^{\infty}t^k f(t)e^{-st}dt
\end{equation}

Taking the limit $s\to 0$:
\begin{equation}
    \begin{split}
        \dfrac{d^k \Tilde{f}(s)}{ds^k}|_{s\to 0}&=(-1)^k\int_0^{\infty}t^k f(t)dt \\
        &=(-1)^k\langle t^k\rangle
    \end{split}
\end{equation}
we find the $k$-th moment of the probability density function $f(t)$:
\begin{equation}
    \langle t^k\rangle=(-1)^k\dfrac{d^k \Tilde{f}(s)}{ds^k}|_{s\to 0}
\end{equation}

Therefore, the $k$-th moment $\langle t_{IJK}^k\rangle$ of the conditional waiting time distribution $\psi_{IJK}(t)$ is:
\begin{equation}
    \langle t_{IJK}^k\rangle=(-1)^k\dfrac{d^k \Tilde{\psi}_{IJK}(s)}{ds^k}|_{s\to 0}
    \label{eq:conditional time moments}
\end{equation}

\section{\label{app:appendixC} Analytical expressions for the 4-state system}
The variables to consider for this system are the mass transition rates $n_{ij}$ and the steady-state probabilities $\pi_i$ for $i,j\in \{1,2,3,4\}$, meaning a total of 16 variables.
Note that $\pi_1$, $\pi_2$, $n_{12}$ and $n_{21}$ are fully observed. Therefore, we are left with 12 variables.
With the following linear constraints, we can immediately reduce the problem to 6 variables.

\subsection{Linear constraints}
We impose probability conservation, mass transition rate conservation in the hidden Markovian states, and mass transition rate conservation between an observed Markovian state and the hidden coarse-grained state.

\subsubsection{Probabilities}
From conservation of the steady-state probability of the Markovian states within the coarse-grained hidden state:
\begin{equation}
    \pi_H=\pi_3+\pi_4
\end{equation}

\subsubsection{Mass conservation at any Markovian state}
We write the mass conservation for one of the hidden states (3 or 4), which for this system, is enough to guaranty the mass conservation for the other hidden state:
\begin{equation}
    n_{13}+n_{23}+n_{43}=n_{31}+n_{32}+n_{34}
\end{equation}

\subsubsection{First-order mass rates}
Here, we require the mass rate conservation of transitions in and out of the hidden state, providing 4 constraint equations:
\begin{equation}
    \begin{split}
        \forall_{i \in \{1,2\}}:n_{iH}=n_{i3}+n_{i4} \\
        \forall_{i \in \{1,2\}}:n_{Hi}=n_{3i}+n_{4i}
    \end{split}
\end{equation}

\subsection{Non-linear constraints}
 The second-order mass transition rates and the conditional waiting times moments can be expressed only as a non-linear function of the optimization problem variables. 
 Here, we show the full derivations of these relations.

\subsubsection{Second-order mass rates}
For this system, as mentioned in the text, we are interested in $n_{1H2}$ and $n_{2H1}$, where the first and the last states are the observed Markovian states.
From equation Eq.~\ref{eq:second-order mass transition rates 1}:
\begin{equation}
    n_{iHj}={\bm n}_{iH}^T\left[\mathbb{I}-{\bm P}_{HH}\right]^{-1} {\bm p}_{H j}
    \label{eq:mass transition through hidden}
\end{equation}
Where
\begin{equation}
    {\bm P}_{HH}=\begin{bmatrix}0 & p_{34} \\ p_{43} & 0\end{bmatrix}
\end{equation}
and
\begin{equation}
    \left[\mathbb{I}-{\bm P}_{HH}\right]^{-1}=\dfrac{1}{1-p_{34}p_{43}}\begin{bmatrix}1 & p_{34} \\ p_{43} & 1\end{bmatrix}
\end{equation}
Plugging into Eq.~\ref{eq:mass transition through hidden}, we have:
\begin{equation}
    \begin{split}
        n_{iHj}&={\bm n}_{iH}^T\left[\mathbb{I}-{\bm P}_{HH}\right]^{-1} {\bm p}_{H j} \\
        &=\begin{bmatrix}n_{i3} & n_{i4}\end{bmatrix}\left(\dfrac{1}{1-p_{34}p_{43}}\begin{bmatrix}1 & p_{34} \\ p_{43} & 1\end{bmatrix}\right)\begin{bmatrix}p_{3j} \\ p_{4j}\end{bmatrix} \\
        &=\dfrac{1}{1-p_{34}p_{43}}\begin{bmatrix}n_{i3} & n_{i4}\end{bmatrix}\begin{bmatrix}p_{3j}+p_{34}p_{4j} \\ p_{4j}+p_{43}p_{3j}\end{bmatrix} \\
        &= \dfrac{n_{i3}(p_{3j}+p_{34}p_{4j})+n_{i4}(p_{4j}+p_{43}p_{3j})}{1-p_{34}p_{43}}
    \end{split}
    \label{eq:second-order mass transition rates - 4state}
\end{equation}
Remember we can express $p_{ij}$ in terms of the mass transition rates (Eq.~\ref{eq:jump probabilities}).

\subsubsection{Conditional waiting time moments}
We calculate the conditional waiting times moments $\langle t_{iHj}^k\rangle$ for $i,j\in \{1,2\}$, in terms of the problem variables.
Based on Eq.~\ref{eq:conditional time moments}, we need to calculate $\Tilde{\psi}_{iHj}^{Normalized}(s)$.

From Eq.~\ref{eq:laplace iJk}:
\begin{equation}
    \Tilde{\psi}_{iHj}(s)=\dfrac{{\bm p}_{iH}^T}{\sum\limits_{h\in\{3,4\}}p_{ih}}\left[\mathbb{I}-\Tilde{{\bm \Psi}}_{HH}(s)\right]^{-1}\Tilde{{\bm \psi}}_{Hj}(s)
    \label{eq:Laplace of iHj}
\end{equation}
Now, we can calculate $\Tilde{{\bm \psi}}_{Hj}(s)$ from Eq.~\ref{eq:laplace vector} and Eq.~\ref{eq:laplace markovian}:
\begin{equation}
    \Tilde{{\bm \psi}}_{Hj}(s)= \begin{bmatrix}\Tilde{\psi}_{3j}(s) \\ \Tilde{\psi}_{4j}(s)\end{bmatrix}=\begin{bmatrix}\dfrac{w_{3j}}{s+\lambda_3} \\ \dfrac{w_{4j}}{s+\lambda_4}\end{bmatrix}
\end{equation}
Given that (Eq.~\ref{eq:laplace matrix} and Eq.~\ref{eq:laplace markovian}):
\begin{equation}
    \begin{split}
        \Tilde{{\bm \Psi}}_{HH}(s)&=\begin{bmatrix}0 & \Tilde{\psi}_{34}(s) \\ \Tilde{\psi}_{43}(s) & 0\end{bmatrix} \\
        &=\begin{bmatrix}0 & \dfrac{w_{34}}{s+\lambda_3} \\ \dfrac{w_{43}}{s+\lambda_4} & 0\end{bmatrix}
    \end{split}
\end{equation}
We can plug into Eq.~\ref{eq:Laplace of iHj}:
\begin{equation}
    \begin{split}
        &\Tilde{\psi}_{iHj}(s)= \\
        &=\dfrac{{\bm p}_{iH}^T}{\sum\limits_{h\in\{3,4\}}p_{ih}}\left[\mathbb{I}-\Tilde{{\bm \Psi}}_{HH}(s)\right]^{-1}\Tilde{{\bm \psi}}_{Hj}(s) \\
        &
        \begin{split}
            =&\dfrac{1}{p_{i3}+p_{i4}}\begin{bmatrix}p_{i3} & p_{i4}\end{bmatrix} \\
            &\left(\left(1-\dfrac{w_{34}w_{43}}{(s+\lambda_3)(s+\lambda_4)}\right)^{-1}\begin{bmatrix}1 & \dfrac{w_{34}}{s+\lambda_3} \\ \dfrac{w_{43}}{s+\lambda_4} & 1\end{bmatrix}\right) \\
            &\begin{bmatrix}\dfrac{w_{3j}}{s+\lambda_3} \\ \dfrac{w_{4j}}{s+\lambda_4}\end{bmatrix}
        \end{split} \\
        &
        \begin{split}
            =&\dfrac{1}{p_{i3}+p_{i4}}\left(1-\dfrac{w_{34}w_{43}}{(s+\lambda_3)(s+\lambda_4)}\right)^{-1} \\
            &\begin{bmatrix}p_{i3} & p_{i4}\end{bmatrix}\begin{bmatrix}\dfrac{w_{3j}}{s+\lambda_3}+\dfrac{w_{34}}{s+\lambda_3}\dfrac{w_{4j}}{s+\lambda_4} \\ \dfrac{w_{4j}}{s+\lambda_4}+\dfrac{w_{43}}{s+\lambda_4}\dfrac{w_{3j}}{s+\lambda_3}\end{bmatrix}
        \end{split} \\
        &
        \begin{split}
            =&\left(1-\dfrac{w_{34}w_{43}}{(s+\lambda_3)(s+\lambda_4)}\right)^{-1} \\
            &\left[\dfrac{p_{i3}}{p_{i3}+p_{i4}}\left(\dfrac{w_{3j}}{s+\lambda_3}+\dfrac{w_{34}}{s+\lambda_3}\dfrac{w_{4j}}{s+\lambda_4}\right) \right. \\
            &\left. +\dfrac{p_{i4}}{p_{i3}+p_{i4}}\left(\dfrac{w_{4j}}{s+\lambda_4}+\dfrac{w_{43}}{s+\lambda_4}\dfrac{w_{3j}}{s+\lambda_3}\right)\right]
        \end{split}
    \end{split}
    \label{eq:laplace 4-state iHj}
\end{equation}

Since the states $i$ and $j$ are Markovian, we just need to normalize this expression in order to get the desired result:
\begin{equation}
    \begin{split}
        &\Tilde{\psi}_{iHj}(s\to 0)= \\
        &
        \begin{split}
            =&\left(1-\dfrac{w_{34}w_{43}}{\lambda_3\lambda_4}\right)^{-1} \\
            &\left[\dfrac{p_{i3}}{p_{i3}+p_{i4}}\left(\dfrac{w_{3j}}{\lambda_3}+\dfrac{w_{34}}{\lambda_3}\dfrac{w_{4j}}{\lambda_4}\right) \right. \\
            &\left. +\vphantom{}\dfrac{p_{i4}}{p_{i3}+p_{i4}}\left(\dfrac{w_{4j}}{\lambda_4}+\dfrac{w_{43}}{\lambda_4}\dfrac{w_{3j}}{\lambda_3}\right)\right]
        \end{split} \\
        &
        \begin{split}
            =\dfrac{p_{i3}\left(p_{3j}+p_{34}p_{4j}\right)+p_{i4}\left(p_{4j}+p_{43}p_{3j}\right)}{\left(p_{i3}+p_{i4}\right)\left(1-p_{34}p_{43}\right)}
        \end{split}
    \end{split}
\end{equation}
Therefore:
\begin{equation}
    \Tilde{\psi}_{iHj}^{Normalized}(s)=\dfrac{\Tilde{\psi}_{iHj}(s)}{\Tilde{\psi}_{iHj}(s\to 0)}
  \label{eq:second-order waiting time - 4state}   
\end{equation}
Finally, we get the moments from Eq.~\ref{eq:conditional time moments}.

In order to get the expressions of the derivatives, we used the package Sympy in Python.

\section{\label{app:appendixD} Analytical expressions for the molecular motor system}
The variables to consider for the molecular motor system are the mass transition rates $n_{u1}$, $n_{u2}$, $n_{d1}$, $n_{d2}$, $n_l$, $n_r$ and the steady-state probabilities $\pi$ and $\pi'$, meaning a total of 8 variables.
With the following linear constraints, we can immediately reduce the problem to 4 variables.

\subsection{Linear constraints}
As in the 4-state system, we impose probability conservation, mass transition rate conservation in the Markovian states, and mass transition rate conservation for the observed transitions $U$ and $D$.

\subsubsection{Probabilities}
From conservation of the steady-state probability of the Markovian states within the coarse-grained states:
\begin{equation}
    \pi + \pi '=\dfrac{1}{3}
\end{equation}

\subsubsection{Mass conservation at any Markovian state}
We write the mass conservation for one of the hidden states (active or passive), which for this system, is enough to guaranty the mass conservation for the other hidden state:
\begin{equation}
    n_r+n_{u2}=n_l+n_{d2}
\end{equation}

\subsubsection{First-order mass rates}
Here, we require the mass rate conservation of transitions in and out of the coarse-grained state, providing 2 constraint equations:
\begin{equation}
    \begin{split}
        n_{U}=n_{u1}+n_{u2} \\
        n_{D}=n_{d1}+n_{d2}
    \end{split}
\end{equation}

\subsection{Non-linear constraints}
Since we have 2 hidden states as in the 4-state system, the results from Appendix~\ref{app:appendixC} can be used here.

\subsubsection{Second-order mass rates}
We use the results for the 4-state system in Eq.~\ref{eq:second-order mass transition rates - 4state}, together with Eq.~\ref{eq:second-order mass transition rates 2}.
For $n_{UU}$, we need to sum over all the mass that goes up from the passive or active state, and then up again only to the passive state:
\begin{equation}
    \begin{split}
        n_{UU} &=\dfrac{n_{u1}(p_{u1}+p_{l}p_{u2})}{1-p_{l}p_{r}} + \dfrac{n_{u2}(p_{u1}+p_{l}p_{u2})}{1-p_{l}p_{r}} \\
        &=\dfrac{(n_{u1}+n_{u2})(p_{u1}+p_{l}p_{u2})}{1-p_{l}p_{r}}
    \end{split}
    \label{eq:nUU}
\end{equation}

For $n_{DD}$, we need to sum over all the mass that goes down only from the passive state, and then down again to the passive or active state:
\begin{equation}
    \begin{split}
        n_{DD} &=\dfrac{n_{d1}p_{d1}+n_{d2}p_{r}p_{d1}}{1-p_{l}p_{r}} + \dfrac{n_{d1}p_{d2}+n_{d2}p_{r}p_{d2}}{1-p_{l}p_{r}} \\
        &=\dfrac{(n_{d1}+n_{d2}p_{r})(p_{d1}+p_{d2})}{1-p_{l}p_{r}}
    \end{split}
    \label{eq:nDD}
\end{equation}

\subsubsection{Conditional waiting time moments}
We account for all of the transitions through a coarse-grained state $i$, and specify in the following calculations the Markovian state before jumping to $i$, and the following Markovian state, after state $i$, 
where $i'$ ($i$) denoted an active (passive) state. For example, $(i-1)\xrightarrow{} (i+1)$ represent two consecutive transitions, $(i-1)\xrightarrow{} i \xrightarrow{} (i+1)$.

Note that a transition upward is only to a passive state, so the previous state (being passive or active) in the first transition does not affect the waiting time. 
Furthermore, a transition downward is only from a passive state.

From Eq.~\ref{eq:laplace IJK}:
\begin{subequations}
    \begin{equation}
        \begin{split}
            \Tilde{\psi}_{UU}(s)&=\dfrac{\pi}{\pi +\pi^{'}}\dfrac{\Tilde{\psi}_{(i-1)\to(i+1)}(s)}{\Tilde{\psi}_{(i-1)\to(i+1)}(s\to 0)} \\
            &+\dfrac{\pi^{'}}{\pi +\pi^{'}}\dfrac{\Tilde{\psi}_{(i-1)'\to(i+1)}(s)}{\Tilde{\psi}_{(i-1)'\to(i+1)}(s\to 0)} \\
            &=\dfrac{\Tilde{\psi}_{(i-1)\to(i+1)}(s)}{\Tilde{\psi}_{(i-1)\to(i+1)}(s\to 0)}
        \end{split}
    \label{eq:timeUU}
    \end{equation}
and similarly:
    \begin{equation}
        \begin{split}
            &\Tilde{\psi}_{UD}(s)= \\
            &=\dfrac{\pi}{\pi +\pi^{'}}\dfrac{\left(\Tilde{\psi}_{(i-1)\to(i-1)}+\Tilde{\psi}_{(i-1)\to(i-1)^{'}}\right)(s)}{\left(\Tilde{\psi}_{(i-1)\to(i-1)}+\Tilde{\psi}_{(i-1)\to(i-1)^{'}}\right)(s\to 0)} \\
            &+\dfrac{\pi^{'}}{\pi +\pi^{'}}\dfrac{\left(\Tilde{\psi}_{(i-1)^{'}\to(i-1)}+\Tilde{\psi}_{(i-1)^{'}\to(i-1)^{'}}\right)(s)}{\left(\Tilde{\psi}_{(i-1)^{'}\to(i-1)}+\Tilde{\psi}_{(i-1)^{'}\to(i-1)^{'}}\right)(s\to 0)} \\
            &=\dfrac{\left(\Tilde{\psi}_{(i-1)\to(i-1)}+\Tilde{\psi}_{(i-1)\to(i-1)^{'}}\right)(s)}{\left(\Tilde{\psi}_{(i-1)\to(i-1)}+\Tilde{\psi}_{(i-1)\to(i-1)^{'}}\right)(s\to 0)}
        \end{split}
        \label{eq:timeUD}
    \end{equation}
    
Moreover:
    \begin{equation}
        \Tilde{\psi}_{DU}(s)=\dfrac{\Tilde{\psi}_{(i+1)\to(i+1)}(s)}{\Tilde{\psi}_{(i+1)\to(i+1)}(s\to 0)}
        \label{eq:timeDU}
    \end{equation}
and:
    \begin{equation}
        \begin{split}
            \Tilde{\psi}_{DD}(s)&=\dfrac{\left(\Tilde{\psi}_{(i+1)\to(i-1)}+\Tilde{\psi}_{(i+1)\to(i-1)^{'}}\right)(s)}{\left(\Tilde{\psi}_{(i+1)\to(i-1)}+\Tilde{\psi}_{(i+1)\to(i-1)^{'}}\right)(s\to 0)} \\
        \end{split}
        \label{eq:timeDD}
    \end{equation}
    \label{eq:laplace general exressions - motor}
\end{subequations}

Now we calculate all the terms in the numerators, using  Eq.~\ref{eq:laplace 4-state iHj} from the 4-state system results:
\begin{subequations}
    \begin{equation}
        \begin{split}
            \Tilde{\psi}_{(i-1)\to(i+1)}(s)=&\left(1-\dfrac{w_{l}w_{r}}{(s+\lambda)(s+\lambda^{'})}\right)^{-1} \\
            &\left(\dfrac{w_{u1}}{s+\lambda}+\dfrac{w_{l}}{s+\lambda}\dfrac{w_{u2}}{s+\lambda^{'}}\right)
        \end{split}
    \end{equation}

    \begin{equation}
        \begin{split}
            &\left(\Tilde{\psi}_{(i-1)\to(i-1)}+\Tilde{\psi}_{(i-1)\to(i-1)^{'}}\right)(s)=\\
            &=\left(1-\dfrac{w_{l}w_{r}}{(s+\lambda)(s+\lambda^{'})}\right)^{-1}\left(\dfrac{w_{d1}}{s+\lambda}+\dfrac{w_{d2}}{s+\lambda}\right) \\
            &=\left(1-\dfrac{w_{l}w_{r}}{(s+\lambda)(s+\lambda^{'})}\right)^{-1}\dfrac{w_{d1}+w_{d2}}{s+\lambda}
        \end{split}
    \end{equation}

    \begin{equation}
        \begin{split}
            &\Tilde{\psi}_{(i+1)\to(i+1)}(s)= \\
            &=\left(1-\dfrac{w_{l}w_{r}}{(s+\lambda)(s+\lambda^{'})}\right)^{-1} \\
            &\left[\dfrac{p_{d1}}{p_{d1}+p_{d2}}\dfrac{w_{u1}}{s+\lambda}+\dfrac{p_{d2}}{p_{d1}+p_{d2}}\dfrac{w_{r}}{s+\lambda^{'}}\dfrac{w_{u1}}{s+\lambda}\right] \\
            &=\left(1-\dfrac{w_{l}w_{r}}{(s+\lambda)(s+\lambda^{'})}\right)^{-1} \\
            &\dfrac{1}{p_{d1}+p_{d2}}\dfrac{w_{u1}}{s+\lambda}\left[p_{d1}+\dfrac{p_{d2}w_{r}}{s+\lambda^{'}}\right]
        \end{split}
    \end{equation}

    \begin{equation}
        \begin{split}
            &\left(\Tilde{\psi}_{(i+1)\to(i-1)}+\Tilde{\psi}_{(i+1)\to(i-1)^{'}}\right)(s)=\\
            &=\left(1-\dfrac{w_{l}w_{r}}{(s+\lambda)(s+\lambda^{'})}\right)^{-1} \\
            &\left[\dfrac{p_{d1}}{p_{d1}+p_{d2}}\dfrac{w_{d1}}{s+\lambda}+\dfrac{p_{d2}}{p_{d1}+p_{d2}}\dfrac{w_{r}}{s+\lambda^{'}}\dfrac{w_{d1}}{s+\lambda}\right.\\
            &\left.+\dfrac{p_{d1}}{p_{d1}+p_{d2}}\dfrac{w_{d2}}{s+\lambda}+\dfrac{p_{d2}}{p_{d1}+p_{d2}}\dfrac{w_{r}}{s+\lambda^{'}}\dfrac{w_{d2}}{s+\lambda}\right] \\
            &=\left(1-\dfrac{w_{l}w_{r}}{(s+\lambda)(s+\lambda^{'})}\right)^{-1} \\
            &\left[\dfrac{p_{d1}}{p_{d1}+p_{d2}}\dfrac{w_{d1}+w_{d2}}{s+\lambda}+\dfrac{p_{d2}}{p_{d1}+p_{d2}}\dfrac{w_{r}}{s+\lambda^{'}}\dfrac{w_{d1}+w_{d2}}{s+\lambda}\right] \\
            &=\left(1-\dfrac{w_{l}w_{r}}{(s+\lambda)(s+\lambda^{'})}\right)^{-1} \\
            &\dfrac{1}{p_{d1}+p_{d2}}\dfrac{w_{d1}+w_{d2}}{s+\lambda}\left[p_{d1}+\dfrac{p_{d2}w_{r}}{s+\lambda^{'}}\right]
        \end{split}
    \end{equation}
    \label{eq:laplace numerators - motor}
\end{subequations}

All of the denominators from Eq.~\ref{eq:laplace general exressions - motor} can be calculated by setting $s\to 0$ in Eq.~\ref{eq:laplace numerators - motor}.
Finally, we get the moments from equation Eq.~\ref{eq:conditional time moments}.

In order to get the expressions of the derivatives, we used the package Sympy in Python.


\clearpage
\bibliographystyle{apsrev4-2}
\bibliography{bib}

\end{document}